\documentclass{pasj00}

\begin{document}
\SetRunningHead{Y. Takeda et al.}{Oxygen and Neon Abundances of 
B-Type Stars}
\Received{2010/06/17}
\Accepted{2010/07/09}

\title{Oxygen and Neon Abundances of B-Type Stars \\
in Comparison with the Sun
\thanks{Based on data obtained at Okayama Astrophysical Observatory
(Okayama, Japan).}
}

%

\author{Yoichi \textsc{Takeda}}
\affil{National Astronomical Observatory of Japan 
2-21-1 Osawa, Mitaka, Tokyo 181-8588}
\email{takeda.yoichi@nao.ac.jp}
\author{Eiji \textsc{Kambe}}
\affil{Okayama Astrophysical Observatory, 
National Astronomical Observatory of Japan\\
Asakuchi, Okayama 719-0232}
\author{Kozo \textsc{Sadakane}}
\affil{Astronomical Institute, Osaka Kyoiku University, 
Asahigaoka, Kashiwara-shi, Osaka 582-8582}
\and 
\author{Seiji \textsc{Masuda}}
\affil{Tokushima Science Museum, Asutamu Land Tokushima, 
Itano-cho, Tokushima 779-0111}

%


\maketitle

\begin{abstract}
To revisit the long-standing problem of possible inconsistency
concerning the oxygen composition in the current galactic gas and 
in the solar atmosphere (i.e., the former being appreciably lower by
$\sim 0.3$~dex) apparently contradicting the galactic chemical evolution, 
we carried out oxygen abundance determinations for 64 mid- through 
late-B stars by using the O~{\sc i} 6156--8 lines while taking into 
account the non-LTE effect, and compared them with the solar O 
abundance established in the same manner.
The resulting mean oxygen abundance was
$\langle A^{\rm O} \rangle = 8.71 (\pm 0.06)$, which means that
 [O/H] (star$-$Sun differential abundance) is $\sim -0.1$, 
the difference being less significant than previously thought. 
Moreover, since the 3D correction may further reduce the reference 
solar oxygen abundance (8.81) by $\sim 0.1$~dex, we conclude that 
the photospheric O abundances of these B stars are almost the same 
as that of the Sun.
We also determined the non-LTE abundances of neon for the sample B stars 
from Ne~{\sc i} 6143/6163 lines to be 
$\langle A^{\rm Ne} \rangle = 8.02 (\pm 0.09)$, leading to the 
Ne-to-O ratio of $\sim 0.2$ consistent with the recent studies. This 
excludes a possibility of considerably high Ne/O ratio once proposed as 
a solution to the confronted solar model problem.
\end{abstract}

%


\section{Introduction}

One of the controversial problems concerned by stellar spectroscopists
over several decades is the inconsistency of CNO abundances
between young early-type stars (e.g., B stars or A--F 
supergiants) and the Sun, where the former should reflect the composition
of the current interstellar gas while the latter represents that of 
rather old galactic gas $\sim 5\times 10^9$~yr ago. That is,
various studies have embarrassingly suggested that the former is appreciably
less than the latter by $\sim 0.3$~dex (see, e.g., Nissen 1993), which is 
just the opposite to what we speculate, because gas metallicity should 
rather increase with an elapse of time as a natural consequence of 
galactic chemical evolution. 

Above all, the discrepancy of oxygen is important, which is (as one of the 
so-called $\alpha$-elements) considered to be synthesized and distributed
essentially by massive stars via type II supernovae. 
Unlike C and N which are known to suffer changes due to evolution-induced 
dredge-up of CN-cycled processed material (e.g., giants or supergiants), 
the photospheric O abundance of any star is believed to generally retain 
the original composition of the gas from which it was formed, 
since the envelope mixing can not penetrate so deep in the interior 
as to salvage the O-burning products according to the standard theory 
of stellar evolution. Do we have to consider the possibility of much 
deeper mixing (i.e., dredge-up of ON-cycle products) due to some unknown 
non-canonical process, as discussed by Luck and Lambert (1985) being faced 
with the result of significant subsolar oxygen abundance by $\sim 0.3$~dex 
in F supergiants and Cepheids? 

Meanwhile, a solution from a different viewpoint has recently been suggested, 
which is based on the argument that the actual solar CNO abundances 
(derived by considering 3D/NLTE corrections along with the most up-to-date 
atomic data) are by $\sim 0.2$~dex lower than those believed so far (cf. 
subsection 4.2 in Asplund et al. 2009). Actually, Przybilla, Nieva, and
Butler's (2008) recent non-LTE study of early B-type stars has demonstrated 
that the abundances agree well with these new solar values.
Though this is surely a straightforward 
explanation (see also Luck \& Lambert 1985), such a ``downward revision'' 
has not been widely accepted yet, because it (in turn) causes a serious 
difficulty in interpreting data of solar seismological observations
within the current framework of solar interior modeling (cf. subsection 4.3 
in Asplund et al. 2009).

Given this confusing situation, we feel it necessary to reinvestigate 
whether or not such a Sun--gas abundance discrepancy really exists, 
based on a well-designed strategy and high-quality observational data,
while paying attention to the following keypoints.

First, which class of stars, being young and representing the composition
of current galactic gas, are the best objects for precisely determining 
the oxygen abundances?  O-type stars are unsuitable for abundance 
studies because of their considerably unstable atmospheres with 
mass loss, while A-type stars tend to show abundance peculiarities 
(typically O-deficiency) in their atmospheres and are unlikely to keep 
their original composition. Regarding high-mass supergiants ranging over 
wide spectral classes (e.g., B--F), sufficient accuracy in abundance
determinations is hard to attain because of the difficulty in modeling
their low-gravity atmospheres (tending to be unstable) as well as
in establishing atmospheric parameters. Accordingly, there is almost 
no better choice than to use B-type main-sequence stars for this purpose.

Second, which spectral lines should then be exploited in order to obtain
differential oxygen abundances between the Sun and B-type stars
as precisely as possible? Admittedly, several elaborate oxygen abundance analyses 
for B-stars have been done so far (e.g., Gies \& Lambert 1992; Kilian 1992; 
Cunha \& Lambert 1994; Korotin et al. 1999; Hempel \& Holweger 2003; 
Przybilla et al. 2008; Niemczura et al. 2009). 
In our opinion, however, the weakpoint of these studies is that, although 
the abundances were established mostly from O~{\sc ii} lines\footnote{
Regarding the analyses of Hempel and Holweger (2003) and Niemczura et al. (2009), 
both of which focused mainly on late-B stars,  O~{\sc i} lines were used.
However, much attention does not seem to have been paid to the reference 
solar oxygen abundance, for which they simply adopted the literature values. 
Besides, the targets of these two studies appear to include not a few 
chemically-peculiar stars, which makes their results comparatively less 
informative, as far as our purpose of investigating the oxygen composition 
for normal B stars is concerned.} 
(particularly seen in the spectra of early-B stars), they were directly 
compared with the solar oxygen abundance (based on O~{\sc i} or [O~{\sc i}] 
or OH lines) simply taken from the literature of solar composition (e.g., 
Anders \& Grevesse 1989; Grevesse et al. 2007; Asplund et al. 2009). 
That is, in order to ensure an accurate abundance determination 
for a star relative to the Sun, a ``differential analysis'' is highly
desirable, which is to carry out an analysis in the same manner for the 
Sun as well as B-stars by using the same line, by which uncertainties in 
$gf$ values (an important source of systematic errors) can be canceled. 
We would here point out based on this standpoint that, among the several 
candidate oxygen lines observable in both B stars as well as in the Sun, 
O~{\sc i} triplet lines at the orange region of 6156--8~$\rm\AA$ are 
most suitable for this purpose, because (1) they are neither too weak 
(reliably measurable for both Sun/B-stars) nor too strong (less 
affected by uncertainties in microturbulence/damping parameters), 
(2) and the non-LTE effect is essentially negligible for the Sun 
(cf. Takeda \& Honda 2005), which means that notorious ambiguities 
in neutral hydrogen collisions (serious source of uncertainties
in non-LTE calculations for late-type stars such as the Sun) are 
irrelevant here.\footnote{This concerns the reason why we do not
invoke the well-known O~{\sc i} triplet lines at 7771--5~$\rm\AA$. 
While these are stronger than O~{\sc i} 6156--8 and thus more easily 
measurable in the spectra of both the Sun and B stars, they suffer 
appreciably large non-LTE corrections. Therefore, uncertainties 
in H~{\sc I} collision rates prevents from precisely establishing 
the solar oxygen abundance if these near-IR triplet lines are to be 
invoked (e.g., appendix 1 in Takeda \& Honda 2005; see also 
Fabbian et al. 2009 or Pereira et al. 2009a, b for the recent progress 
in this field).}

Thus, the primary motivation of this study is to determine the oxygen 
abundances of selected B-type main-sequence stars based on the
O~{\sc i} 6156--8 lines and compare them with the solar abundance 
in order to see if there is any difference between these, while making 
use of our past experiences in O abundance determinations with these 
triplet lines for late-B through F supergiants or late-B/A dwarfs or 
FGK dwarfs including the Sun (cf. Takeda \& Takada-Hidai 1998; 
Takeda et al. 1999; Takeda \& Honda 2005).

In addition, as a  by-product of the analysis, we decided to 
study the neon abundances of B-type stars, since Ne~{\sc i} 6143/6163 
lines suitable for abundance determinations exist in the
targeted spectral region. Though comparatively minor interest seems to 
have paid to Ne and only a few abundance studies were available 
in the past (e.g., Auer \& Mihalas 1973; Dworetsky \& Budaj 2000;
Hempel \& Holweger 2003), this $\alpha$ element (similar to oxygen) 
got particular attention recently and elaborate abundance studies 
have been published (Cunha et al. 2006; Morel \& Butler 2008) 
because a possibility of considerably higher Ne/O ratio (than 
has ever been thought) was proposed as a key to resolve the serious 
solar model problem encountered for the case of downward revision
of solar CNO abundances mentioned above. Though recent studies tend to 
be against this possibility (see also subsection 4.3 in Asplund et al. 2009), 
we intend to check it independently by ourselves, which constitutes 
the second aim of the present investigation.

\section{Observational Data}

The targets of this study are the selected 64 mid-though-late B-type 
stars mostly with spectral classes of B2--B9 and luminosity classes of 
III--V (cf. table 1), which are apparently bright ($V \sim$~4--6.5) 
as well as sharp-lined ($v_{\rm e}\sin i \ltsim 30$~km~s$^{-1}$ in 
Abt \& Morrell's 1995 catalogue). These program stars are plotted on 
the $\log L$ vs. $\log T_{\rm eff}$ diagram in figure 1, where 
theoretical evolutionary tracks corresponding to different stellar 
masses are also depicted. We can see from this figure that the masses 
of our sample stars are in the range between $\sim 3 M_{\odot}$ and 
$\sim 9 M_{\odot}$.

\setcounter{figure}{0}
\begin{figure}
  \begin{center}
    \FigureFile(80mm,80mm){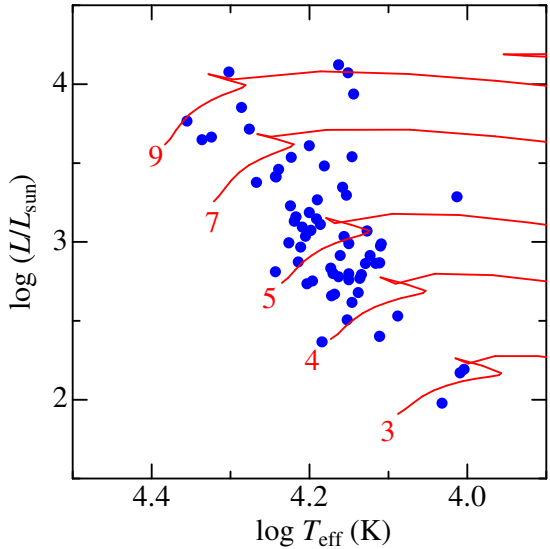}
  \end{center}
\caption{Plots of 64 program stars on the theoretical HR diagram
($\log (L/L_{\odot})$ vs. $\log T_{\rm eff}$), where the bolometric 
luminosity ($L$) was evaluated from the apparent visual magnitude with 
the help of Hipparcos parallax (ESA 1997), Arenou et al.'s (1992)
interstellar extinction correction, and  Flower's (1996) bolometric
correction. Theoretical evolutionary tracks corresponding to 
the solar metallicity computed by Lejeune and Schaerer (2001) for 
five different initial masses (3, 4, 5, 7, and 9 $M_{\odot}$) 
are also depicted for comparison.
}
\end{figure}

The observations were carried out on 2006 October 17--23 by using 
the HIgh-Dispersion Echelle Spectrograph (HIDES; Izumiura 1999) 
at the coud\'{e} focus of the 188~cm reflector of Okayama Astrophysical 
Observatory (OAO). Equipped with a 4K$\times$2K CCD detector at 
the camera focus, the HIDES spectrograph enabled us to obtain an 
echellogram covering a wavelength range of 5600--6800~$\rm\AA$ 
with a resolving power of $R \sim 70000$ (case for the normal slit 
width of 200~$\mu$m) in the mode of red cross-disperser.

The reduction of the spectra (bias subtraction, flat-fielding, 
scattered-light subtraction, spectrum extraction, wavelength 
calibration, and continuum normalization) was performed by using 
the ``echelle'' package of the software IRAF\footnote{
IRAF is distributed by the National Optical Astronomy Observatories,
which is operated by the Association of Universities for Research
in Astronomy, Inc. under cooperative agreement with the National 
Science Foundation.} in a standard manner. 
For most of the targets, we could accomplish sufficiently high S/N 
ratio of several hundreds. 

\section{Atmospheric Parameters}

The effective temperature ($T_{\rm eff}$) and the surface gravity 
($\log g$) of each program star were determined from the colors of 
Str\"{o}mgren's $uvby\beta$ photometric system with the help of 
Napiwotzki, Sc\"{o}nberner, and Wenske's (1993) {\tt uvbybetanew}
program,\footnote{
$\langle$http://www.astro.le.ac.uk/\~{}rn38/uvbybeta.html$\rangle$.} 
which is based on Moon's (1985) {\tt UVBYBETA} (estimation of intrinsic 
colors corrected for interstellar reddening) and
{\tt TEFFLOGG} ($T_{\rm eff}$/$\log g$ determination from dereddened
colors) programs. The observational data of
$b-y$, $c_{1}$, $m_{1}$, and $\beta$ were taken from Hauck and 
Mermilliod (1998) via the SIMBAD database.
The resulting $T_{\rm eff}$ and $\log g$ are summarized in table 1.
Their typical errors may be estimated as $\sim 3\%$ in 
$T_{\rm eff}$ and $\sim 0.2$~dex in $\log g$ for the present
case of mid-through-late B stars, according to Napiwotzki et al. 
(1993; cf. their section 5).
The model atmosphere for each star was then constructed
by two-dimensionally interpolating Kurucz's (1993) ATLAS9 
model grid in terms of $T_{\rm eff}$ and $\log g$, where
we exclusively applied the solar-metallicity models. 

Regarding the microturbulence ($\xi$), which is necessary for abundance
determinations, we could not establish this parameter based on 
our spectra because the number of strong lines was not enough.
According to Lyubimkov, Rostopchin, and Lambert (2004),
the range of $\xi$ values is $\sim$~0--5~km~s$^{-1}$
for B-type main-sequence stars with masses of 4--11~$M_{\odot}$
(cf. their figures 13 and 14). Therefore, we tentatively assumed 
$\xi$ = 3~km~s$^{-1}$ with possible uncertainties of $\pm 2$~km~s$^{-1}$
for all our sample stars. Fortunately, as shown in subsection 4.3 
(cf. figures 3f and 4f), the resulting oxygen and neon abundances are 
not significantly affected by these ambiguities in $\xi$, thanks to 
the moderate strengths of the relevant spectral lines as well as 
to the large thermal velocity (which makes the role of microturbulence 
comparatively insignificant) because of their being light atoms 
(atomic numbers are 8 and 10).

\section{Abundance Determinations}

\subsection{Non-LTE Calculations}

The non-LTE calculations for oxygen were performed in the same manner 
as in Takeda and Honda's (2005) analysis on oxygen lines of solar-type stars
by using an O~{\sc i} atomic model consisting of 87 terms and 277 radiative 
transitions.\footnote{Regarding the work of Takeda and Takada-Hidai (1998)
and Takeda et al. (1999), which also focused on the analysis of 
O~{\sc i} 6156--8 lines in late B--F supergiants and late B--A
dwarfs, respectively, a slightly different O~{\sc i} atomic model 
comprising 86 terms and 294 radiative transitions (Takeda 1992, 1997)
was used. However, as far as the relevant O~{\sc i} 6156--8
high-excitation lines are concerned, both atomic models give essentially
the same results.} See Takeda (2003) and the references therein for 
more computational details. 

As for the statistical-equilibrium calculations for Ne, we newly constructed 
an atomic model of Ne~{\sc i} comprising 94 terms (including up to 
$2p^{5} 10d'$ with $E = 173609$~cm$^{-1}$) and 1034 radiative transitions, 
based on Kurucz and Bell's (1995) atomic data.
Regarding the photoionization cross sections, we adopted the data 
provided by the Opacity Project via the TOPbase system (Cunto, Mendoza 1992) 
for the lowest 46 terms, while the hydrogenic approximation was assumed 
for the remaining terms. 
Regarding the collisional rates, we followed the recipe adopted in 
subsubsection 3.1.3 of Takeda (1991), which is largely based on the 
treatment of Auer and Mihalas (1973). Note that neutral-hydrogen collisions 
(though included formally) are essentially negligible compared to 
electron collisions in the present case of B-type stars.

We carried out non-LTE calculations for O as well as for Ne on a grid of 
36 (=9$\times$4) model atmospheres resulting from combinations of nine 
$T_{\rm eff}$ values (9000, 10000, 12000, 14000, 16000, 18000, 20000, 22000, 
24000~K) and four $\log g$ values (3.0, 3.5, 4.0, 4.5) with Anders and 
Grevesse's (1989) solar oxygen and neon abundances of 8.93 (O) and 8.09 (Ne) 
(being consistent with the ATLAS9 model atmospheres we use in this study),
so that we can obtain the depth-dependent non-LTE departure coefficients
for any star by interpolating this grid. 

\subsection{Synthetic Spectrum Fitting}

Now that the non-LTE departure coefficients for O~{\sc i} and Ne~{\sc i}
lines relevant for each star are available, with which the non-LTE 
theoretical spectrum of O~{\sc i} 6156--8 and Ne~{\sc i} 6143/6163 lines 
can be computed, we carried out spectrum-synthesis analyses by applying 
Takeda's (1995) automatic-fitting procedure to the 6140--6165~$\rm\AA$ region 
while regarding $A^{\rm O}$ and $A^{\rm Ne}$ (as well as $v_{\rm e}\sin i$ 
and radial velocity) as adjustable parameters
to be established.\footnote{The abundances of C ($A^{\rm C}$) or 
Fe ($A^{\rm Fe}$) were also treated as variables in some cases where 
features of these lines were appreciable.} 
The adopted atomic data of the relevant O~{\sc i} and Ne~{\sc i} 
lines are presented in table 2.
How the theoretical spectrum for the converged solutions fits well 
with the observed spectrum is displayed in figure 2 and the resulting 
abundances ($A^{\rm O}$ and $A^{\rm Ne}$) are given in table 1.

\subsection{Equivalent Widths and Abundance Uncertainties}

While the non-LTE synthetic spectrum fitting directly yields
the final abundance solution, this approach is not necessarily
suitable when one wants to evaluate the extent of non-LTE corrections or 
to study the abundance sensitivity to changing the atmospheric parameters 
(i.e., it is tedious to repeat the fitting process again and 
again for different assumptions or different atmospheric parameters).
Therefore, with the help of Kurucz's (1993) WIDTH9 program (which had 
been considerably modified in various respects; e.g., inclusion of
non-LTE effects while multiplying the line opacity as well as the line 
source function by appropriate factors computed from departure coefficients, 
treatment of total equivalent width for multi-component blended lines by
integrating the synthesized spectrum; etc.), we computed 
the equivalent widths for the O~{\sc i} 6156--8 triplet 
as a whole\footnote{We here computed total equivalent width
for the whole triplet (i.e., not the equivalent widths for the three 
individual features) in order to maintain consistency (or to enable 
direct comparison) with Takeda and Takada-Hidai (1998) and Takeda et al. 
(1999), where the triplet lines were occasionally merged and hard to 
resolve due to considerable line broadening.} ($W_{6156-8}^{\rm O}$), 
Ne~{\sc i} 6143 line ($W_{6143}^{\rm Ne}$), and Ne~{\sc i} 6163 line 
($W_{6163}^{\rm Ne}$) ``inversely'' from the abundance solutions 
(resulting from non-LTE spectrum synthesis) along with the adopted 
atmospheric model/parameters, since they are much easier to handle.
Based on such evaluated $W$ values, the non-LTE as well as LTE abundances 
were freshly computed to derive the non-LTE correction ($\Delta$).
These $W$ and $\Delta$ values are also given in table 1.

We then estimated the uncertainties in $A^{\rm O}$ and $A^{\rm Ne}$
by repeating the analysis on the $W$ values while perturbing
the standard atmospheric parameters interchangeably by $\pm 3\%$ 
in $T_{\rm eff}^{\rm std}$, $\pm 0.2$~dex in $\log g^{\rm std}$, 
and $\pm 2$~km~s$^{-1}$ in $\xi^{\rm std}$ (which are the typical 
uncertainties of the parameters we adopted; cf. section 3). 
Figures 3 (O) and 4 (Ne) graphically show the resulting abundances,
equivalent widths, non-LTE corrections, and abundance variations
in response to parameter changes as functions of $T_{\rm eff}$.

\setcounter{figure}{2}
\begin{figure}
  \begin{center}
    \FigureFile(80mm,140mm){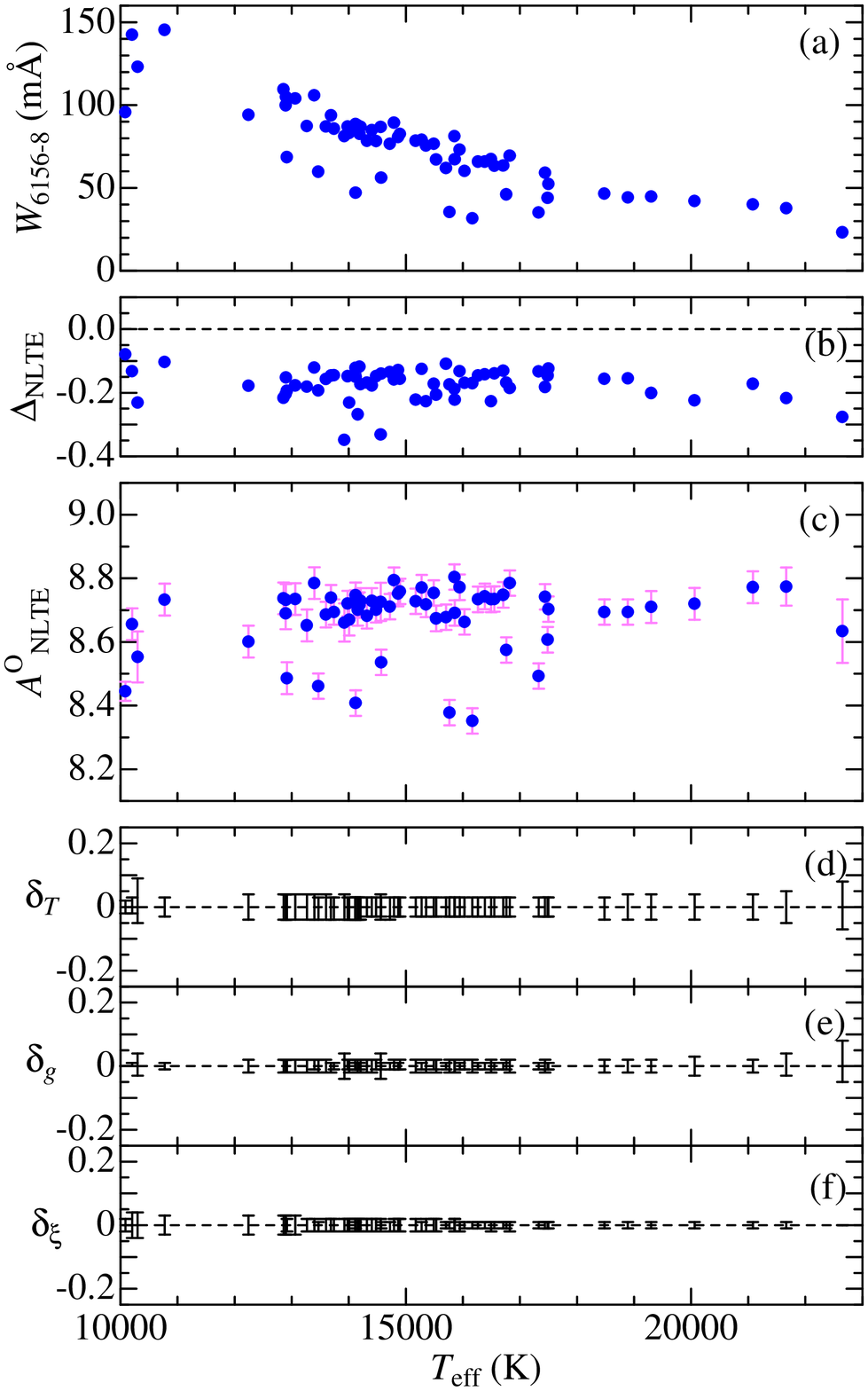}
  \end{center}
\caption{Oxygen abundances derived from synthetic spectrum 
fitting, along with the abundance-related quantities specific 
to the O~{\sc i} 6156--8 triplet, plotted against $T_{\rm eff}$. 
(a) $W_{6156-8}$ (total equivalent width), 
(b) $\Delta_{\rm NLTE}$ (non-LTE correction),
(c) $A_{\rm NLTE}^{\rm O}$ (non-LTE oxygen abundance;
where the indicated error bars represent the root-sum-square 
of $\delta_{T} [\equiv (|\delta_{T+}| + |\delta_{T-}|)/2]$, 
$\delta_{g} [\equiv (|\delta_{g+}| + |\delta_{g-}|)/2]$, 
and $\delta_{\xi} [\equiv (|\delta_{\xi +}| + |\delta_{\xi -}|)/2]$, 
(d) $\delta_{T+}$ and $\delta_{T-}$ (abundance variations 
in response to $T_{\rm eff}$ changes of $+3 \%$ and $-3 \%$), 
(e) $\delta_{g+}$ and $\delta_{g-}$ (abundance variations 
in response to $\log g$ changes of $+0.2$~dex and $-0.2$~dex), 
and (f) $\delta_{\xi +}$ and $\delta_{\xi -}$ (abundance 
variations in response to perturbing the standard $\xi$ value
by $\pm 2$~km~s$^{-1}$; i.e., changing to 5~km~s$^{-1}$ and 1~km~s$^{-1}$).
The signs of $\delta$'s are $\delta_{T+}>0$, $\delta_{T-}<0$,
$\delta_{g+}<0$, $\delta_{g-}>0$, $\delta_{\xi +}<0$, and
$\delta_{\xi -}>0$.
}
\end{figure}

\setcounter{figure}{3}
\begin{figure}
  \begin{center}
    \FigureFile(80mm,140mm){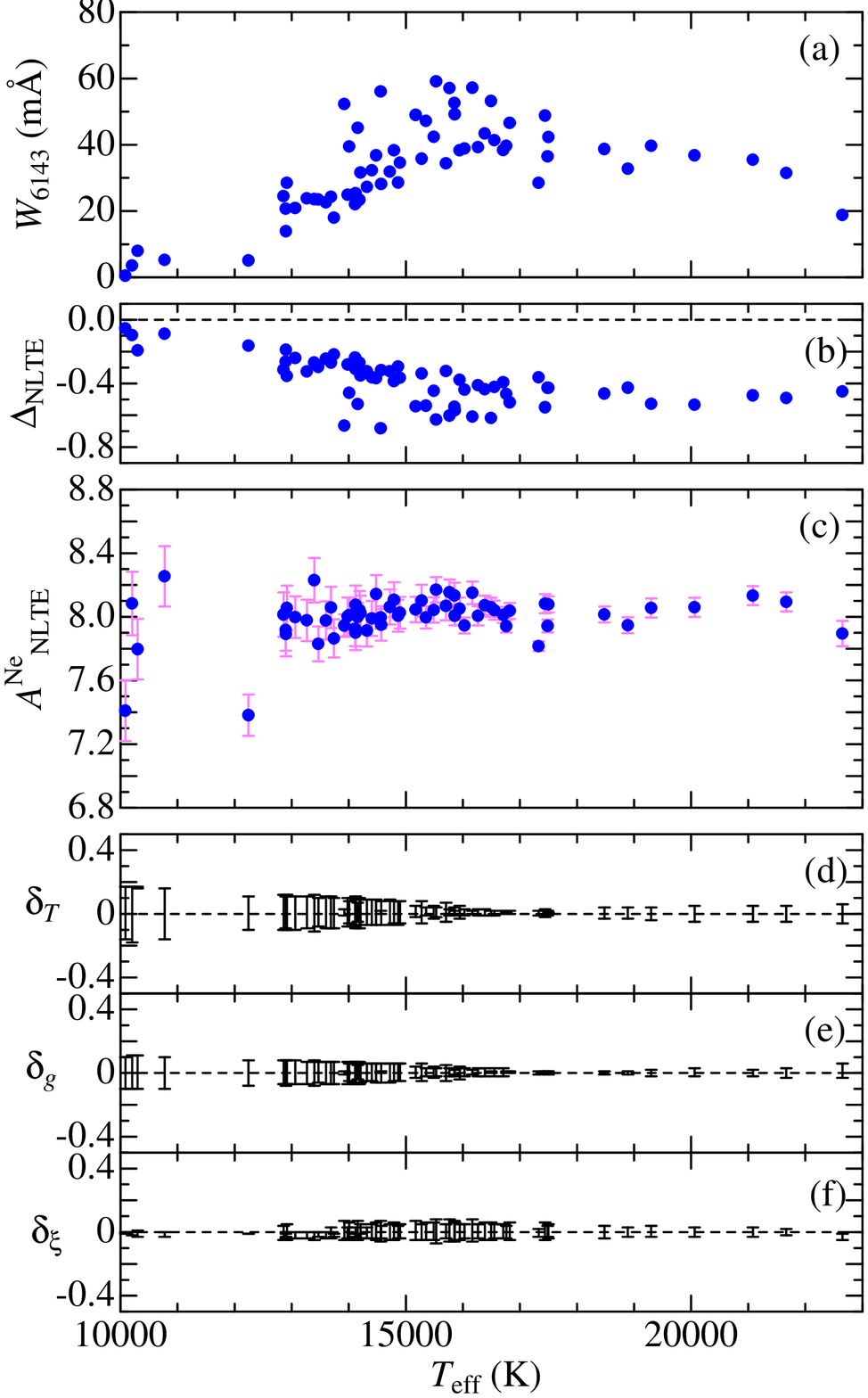}
  \end{center}
\caption{
Neon abundances derived from synthetic spectrum 
fitting, along with the abundance-related quantities specific 
to the Ne~6143 line (the stronger Ne line of the two), plotted 
against $T_{\rm eff}$.
Some cautions should be made regarding the signs of $\delta$'s:
while $\delta_{\xi +}<0$ and $\delta_{\xi -}>0$ generally hold
for all the $T_{\rm eff}$ range, the signs of Ne abundance 
variations in response to changing $T_{\rm eff}$ and $\log g$ 
are inversed around some critical $T_{\rm eff}$ ($\sim$~17000--18000~K);
that is, $\delta_{T+}<0$ ($\delta_{T-}>0$) and $\delta_{g+}>0$ 
($\delta_{g-}<0$) on the lower-$T_{\rm eff}$ side, and vice versa
on the higher-$T_{\rm eff}$ side. Otherwise, the same as in figure 3.
}
\end{figure}

\section{Discussion}

\subsection{Oxygen: Consistency with the Solar Composition}

We can now compare the resulting oxygen abundances of B-type stars
with the solar oxygen abundance ($A_{\odot}^{\rm O} = 8.81$)
derived by Takeda and Honda (2005) in the same manner as we did
in this study (i.e., spectrum fitting of O~{\sc i} 6156--8 lines). 
The histogram of star$-$Sun differential abundances ([O/H] 
$\equiv A^{\rm O} - 8.81$) for the 64 stars is depicted in figure 5,
where the result for late B through F supergiants derived by
Takeda and Takada-Hidai (1998) is also shown.\footnote{Though [O/H] 
values were computed with Anders and Grevesse's (1989) solar abundance 
of 8.93 in that paper, they have been rescaled by using the solar 
abundance of 8.81 adopted here.}
As manifestly seen in figure 3c, some $\sim$~5--10 stars at 
$T_{\rm eff} \ltsim 17000$~K exhibit appreciable discrepancies
from the main trend (i.e., O-deficiency by several tenths dex),
which means that chemically peculiar stars are included in late-B
stars of lower $T_{\rm eff}$. These outliers are recognized
in figure 5 as a long tail at [O/H] $\ltsim -0.3$.
Thus, excluding those 7 stars with $A^{\rm O} < 8.51$ or [O/H] $ < -0.3$
(HD~223229, 185330, 30122, 49606, 53244, 23408, 181470),
we conclude by averaging the results for the remaining 57 stars that
$\langle A^{\rm O} \rangle = 8.71 (\pm 0.06)$ (the value following 
$\pm$ is the standard deviation) or 
$\langle$[O/H]$\rangle = -0.10 (\pm 0.06$).
This leads to a consequence that the oxygen abundance 
deficit of B stars relative to the Sun has been appreciably reduced
down to $\sim 0.1$~dex, as compared to the previously claimed value 
of $\sim 0.3$~dex. 

\setcounter{figure}{4}
\begin{figure}
  \begin{center}
    \FigureFile(80mm,80mm){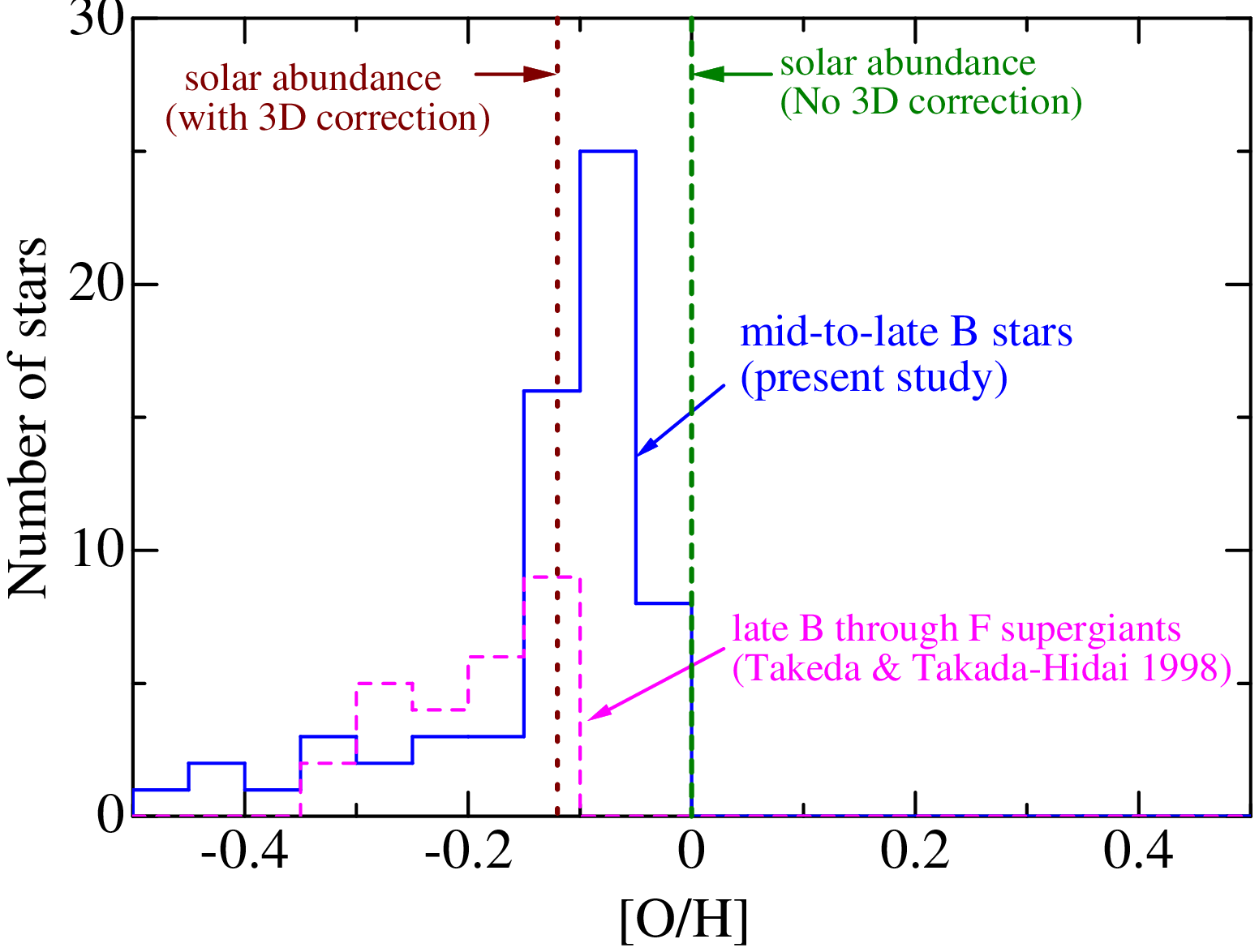}
  \end{center}
\caption{Histogram of oxygen abundances relative 
to the Sun ([O/H] $\equiv A_{\rm NLTE}^{\rm O} - 8.81$),
where the solar oxygen abundance derived by Takeda and Honda (2005) 
with Kurucz's (1993) 1D model is here adopted as the reference abundance 
as shown by the vertical dashed (green) line, while the position for 
the modified solar abundance corrected for the 3D correction 
($-0.12$~dex) is also marked by the vertical dotted (brown) line.
These indicated positions of the reference solar oxygen abundance would be
shifted leftward by $-0.06$~dex (no 3D) and $-0.08$~dex (with 3D), if 
Holweger and M\"{u}ller's (1974) semi-empirical solar photospheric model 
is used instead of Kurucz's (1993) model atmosphere for the Sun. 
Solid (blue) line $\cdots$ mid-to-late B-type stars 
derived in this study. Dashed (red line)
$\cdots$ late B through F supergiants based on 
the results of Takeda and Takada-Hidai (1998). 
(See subsection 5.1 for a remark concerning the comparison of the current 
photospheric abundances between the Sun and B-type stars.)
}
\end{figure}

Moreover, there is a possibility that even this small gap may almost
vanish in the end. The reference solar oxygen abundance (8.81) 
we adopted here is the value derived by Takeda and Honda (2005) 
by a non-LTE analysis of O~{\sc i} 6156--8 lines in Moon (solar flux) 
spectra using Kurucz's (1993) ATLAS9 solar (1D) model atmosphere, 
just similar to the present study of B stars. While the non-LTE effect 
turned out to be negligible for these triplet lines in the solar case, 
no consideration was made for the 3D effect due to atmospheric 
inhomogeneities caused by convective granular motions.
We should alert to the correction for this effect (3D correction) 
to be applied to the solar oxygen abundance derived from classical 
1D model atmospheres, though it should be irrelevant for B-type stars. 
As a matter of fact, an appreciable amount of this correction seems 
probable. According to Caffau et al. (2008), the 3D correction
for the 1D oxygen abundance determined from the O~{\sc i} 6158 
line with Holweger and M\"{u}ller's (1974) solar model is
$\Delta$(3D$-$1D$_{\rm HM}$) = $-0.14$, while that with Kurucz's ATLAS
solar model is $\Delta$(3D$-$1D$_{\rm KU}$) = $-0.12$ (cf. their table 3
and subsubsection 5.3.3). Asplund et al. (2004; cf. their table 3) 
derived almost the same result ($-0.15$) for $\Delta$(3D$-$1D$_{\rm HM}$). 
Since Caffau et el.'s (2008) $\Delta$(3D$-$1D$_{\rm KU}$) correction 
should be relevant for Takeda and Honda's (2005) case, our reference
solar oxygen abundance could be lowered by $-0.12$~dex to be 8.69,
which is in remarkable agreement with the average (8.71) of B-stars 
mentioned above. Consequently, we may state that ``the oxygen abundances 
of young mid-through-late B stars, which are considered to keep the 
composition of current galactic gas, are consistent with the solar 
photospheric oxygen abundance without any significant discrepancy.''

Yet, from a quantitative point of view, some care should be made 
in comparing the current photospheric oxygen abundances of the Sun 
and B stars, since the composition of the galactic gas as well as 
that of the solar photosphere (both of which must have been the same 
at the birth of the Sun) have suffered slight changes over these 
$\sim 5\times 10^{9}$~yr due to two processes:\\
--- First, a small enrichment in the galactic gas ($\sim 0.04$~dex 
for the case of O and Ne; see, e.g., Chiappini et al. 2003) is expected
as a result of the galactic chemical evolution (for the present case of 
$\alpha$ elements, they are synthesized and distributed by massive 
stars via type II supernovae). \\
--- Second, as a consequence of the diffusion process in the envelope
of the Sun, the current solar photospheric abundance is supposed have 
suffered a marginal decrease ($\sim 0.04$~dex; cf. subsection 3.11 in 
Asplund et al. 2009) as compared to the bulk solar abundance (i.e., 
original composition).\\
Accordingly, from a purely theoretical prediction, the oxygen 
composition of B-type stars (reasonably regarded as equivalent to that 
in the current interstellar gas) would be somewhat higher by 
$\sim 0.1$~dex (0.04 + 0.04) than that in the solar photosphere. 
Therefore, we should bear in mind that too much meaning should not 
be put to the apparently very good agreement (8.69 and 8.71) we obtained 
here, and that the presently accomplished consistency is nothing but on the 
order of $\ltsim 0.1$~dex. There may be some other factors to be considered 
if a much better quantitative consistency is to be pursued.\footnote{
As a possibility, some ambiguity may still exist in the reference solar
oxygen abundance, depending on the solar photospheric model, for which 
we adopted Takeda and Honda's (2005) result derived by using 
Kurucz's (1993) ATLAS9 model (KU) for the solar atmosphere. 
In order to check this point, we carried out a spectrum-fitting analysis 
on the O~{\sc i} 6156--8 feature in the Moon spectrum in the same way 
as done by Takeda and Honda (2005) but with Holweger and M\"{u}ller's 
(1974) semi-empirical solar model (HM), and obtained a slightly lower 
solar oxygen abundance by 0.06~dex (see also table 3 of Takeda 1994). 
If we further consider the difference of 0.02~dex in the extent of 
(negative) 3D correction mentioned above [$\Delta$(3D$-$1D$_{\rm HM}$) 
$-$ $\Delta$(3D$-$1D$_{\rm KU}$) = $(-0.14)-(-0.12)$], the 3D solar oxygen 
abundance corresponding to HM model would be 8.61 (i.e., $8.69 - 0.06 - 0.02$). 
Interestingly, this is by 0.1~dex smaller than the mean abundance of B-type 
stars, and is more consistent with the theoretical expectation.}

Finally, the overview display for the oxygen abundances of various types
of stars derived from O~{\sc i} 6156--8 lines in the same manner
is presented in figure 6, where the results of this study are 
combined with those of Takeda and Takada-Hidai (1998), Takeda et al. 
(1999), and Takeda and Honda (2005). We can see that the oxygen abundances
of B-type stars (this study) and late B--F supergiants (Takeda \& Takada-Hidai
1998) are almost within $\pm \ltsim$~0.1--0.2~dex around the solar abundance,
especially when the 3D-correction is taken into account.

\setcounter{figure}{5}
\begin{figure}
  \begin{center}
    \FigureFile(80mm,80mm){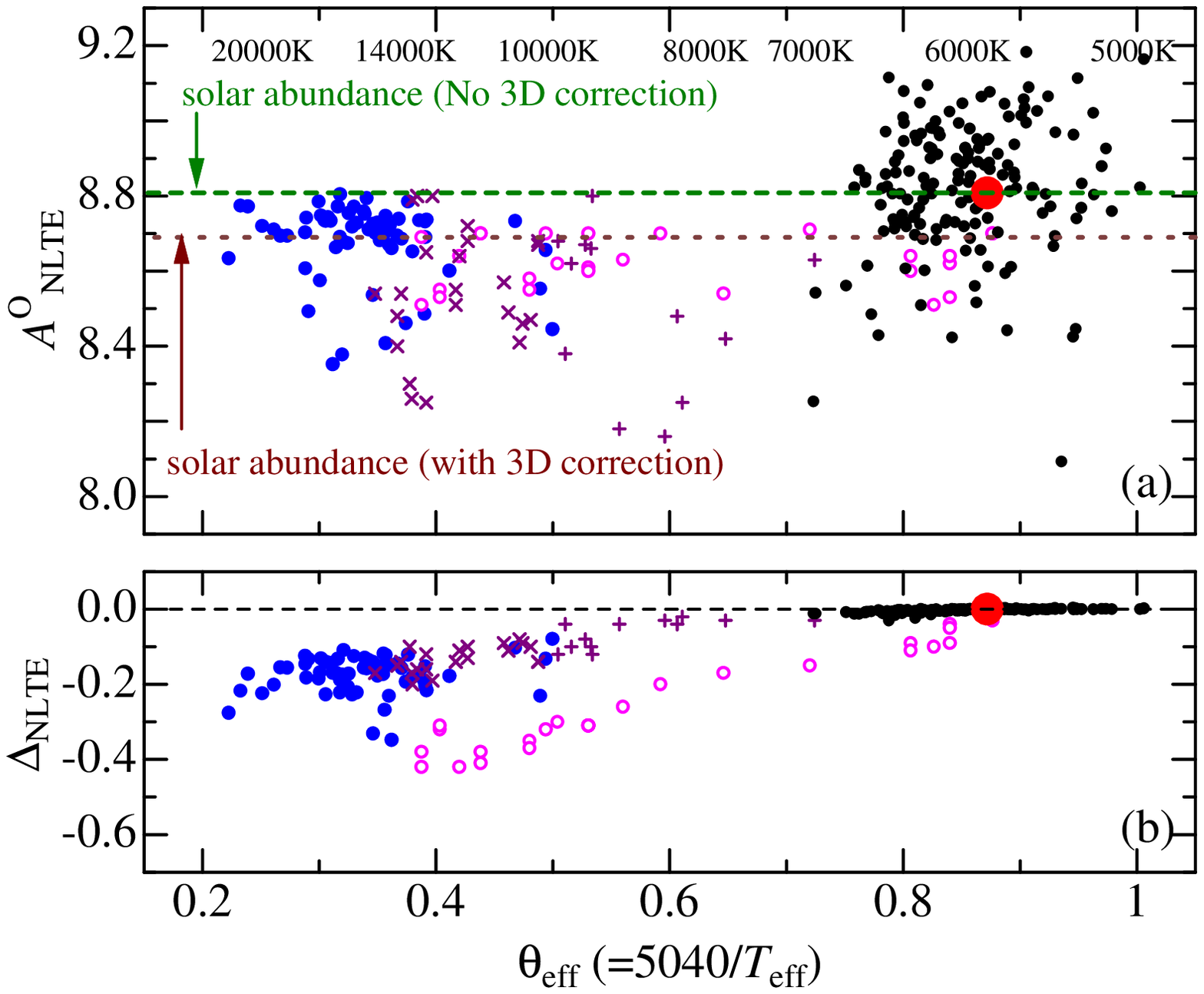}
  \end{center}
\caption{Non-LTE oxygen abundances ($A_{\rm NLTE}^{\rm O}$; 
upper panel (a)) along with the corresponding non-LTE 
corrections ($\Delta_{\rm NLTE}$; lower panel (b)), 
which were derived for various class of stars in the same manner
as this study (i.e., profile fitting on the O~{\sc i} 6156--8 triplet), 
plotted against $\theta_{\rm eff} (\equiv 5040/T_{\rm eff})$.
Filled circles (blue) $\cdots$ mid-to-late B stars (this study),
open circles (pink) $\cdots$ late B through F supergiants 
(Takeda \& Takada-Hidai 1998), St. Andrew's crosses ($\times$; purple)
$\cdots$ non-variable late-B dwarfs (Takeda et al. 1999), 
Greek crosses (+; purple) $\cdots$ non-variable A dwarfs 
(Takeda et al. 1999), and dots (black) $\cdots$ FGK dwarfs 
(Takeda \& Honda 2005). The solar abundance derived by Takeda
and Honda (2005) with Kurucz's (1993) 1D model atmosphere is 
indicated by a big (red) filled circle along with the horizontal 
dashed (green) line (panel (a)). Another solar oxygen abundance 
corrected for 3D effect of $-0.12$~dex according to Caffau et al. 
(2008) is also depicted in panel (a) by the horizontal dotted 
(brown) line.
}
\end{figure}

\subsection{Neon: Confirmation of Recent Results}

Figure 4c indicates that the distribution of neon abundances derived 
for the program stars are well homogeneous, which can also be 
recognized by the histogram of $A^{\rm Ne}$ displayed in figure 7.
Yet, only two stars (HD~178065 and HD~181470), both of which are
near to the low-$T_{\rm eff}$ end ($T_{\rm eff} \sim$~10000--12000~K)
among the targets, show a marked deficit (by $\sim -0.6$~dex) 
compared to others. This may reflect the tendency of Ne-deficiency 
presumably due to gravitational settling in late-B chemically peculiar 
stars (such as HgMn stars) reported by Dworetsky and Budaj (2000). 
Excluding these two stars, we obtain 
$\langle A^{\rm Ne} \rangle = 8.02 (\pm 0.09)$.
This neon abundance for mid-to-late B stars agrees fairly well with 
the recent result of Morel and Butler (2008), who concluded 
$7.97 \pm 0.07$ for 18 nearby early-B stars based on their non-LTE 
analysis on Ne~{\sc i} and Ne~{\sc ii} lines.
It is worth mentioning that their value becomes $8.03 \pm 0.08$ (i.e.,
almost in perfect agreement) when only Ne~{\sc i} lines were used 
(cf. their subsection 5.1) as we did. 
Our result is also in accord with the value of $8.08 \pm 0.03$ derived by 
Przybilla, Nieva, and Butler (2008) for six B-type stars within the error bar.
Similarly, a reasonable consistency is seen with other previously published
work of Ne abundances of B-type stars (Kilian 1994; Cunha et al. 2006;
Hempel \& Holweger 2003; Dworetsky \& Budaj 2000; Sigut 1999), though
these studies suggested a marginally high (by 0.1~dex) neon abundance
of $A^{\rm Ne} \sim 8.1$ than ours (see subsection 5.1 of 
Morel \& Butler 2008 for an extensive discussion). 

\setcounter{figure}{6}
\begin{figure}
  \begin{center}
    \FigureFile(80mm,80mm){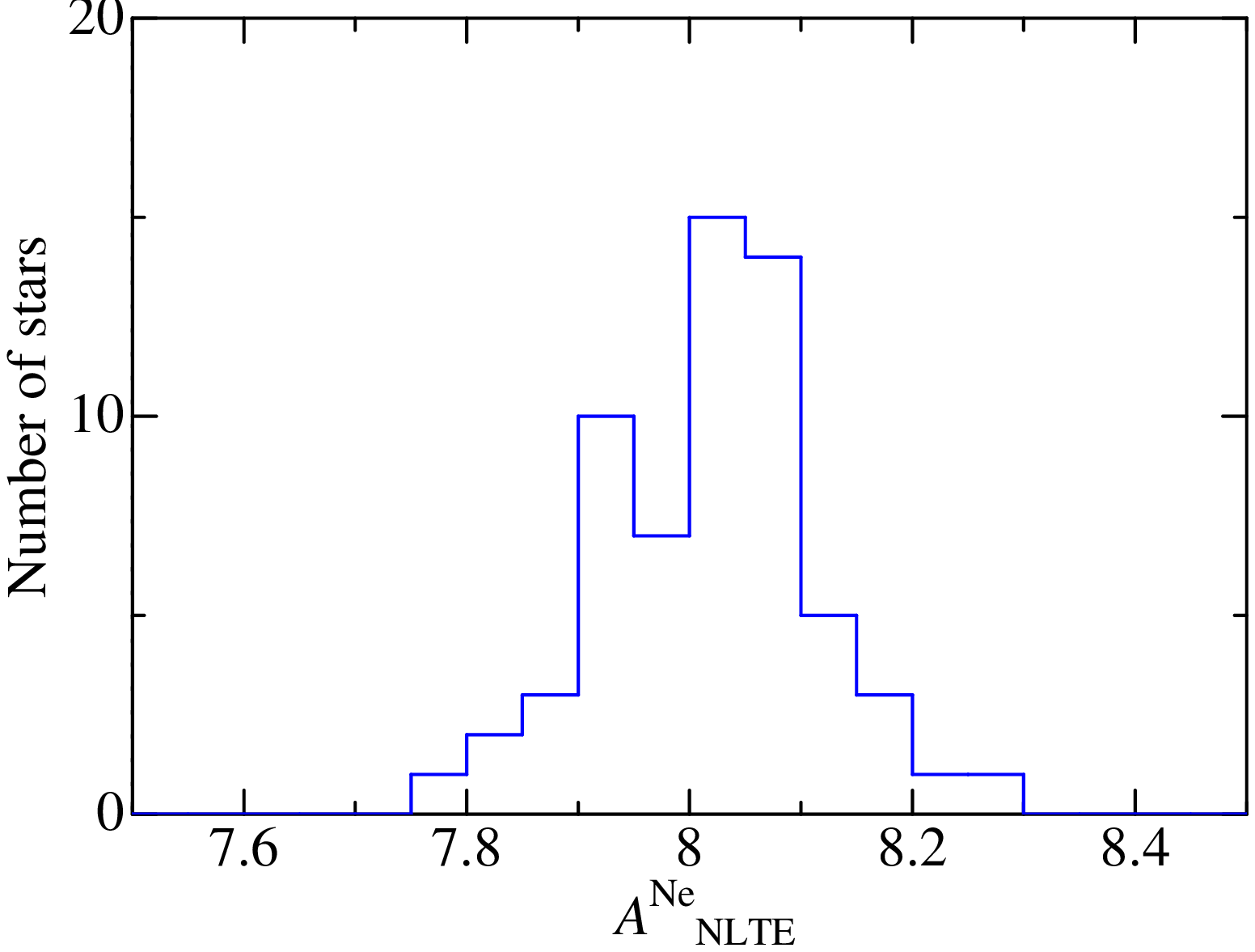}
  \end{center}
\caption{Histogram of neon abundances derived in this study
($A_{\rm NLTE}^{\rm Ne}$). Markedly low Ne abundances ($\sim 7.4$) 
exhibited by the two stars (HD~178065 and HD~181470) are
outside of the range of the abscissa.
}
\end{figure}

Regarding the connection to the solar Ne abundance, especially
with regard to a possibility of considerably high Ne/O ratio 
($\sim 0.5$) which might resolve the serious solar model
problem encountered for the case of downward revision in the 
CNO abundances by $\sim 0.2$~dex (i.e., significant discrepancy
between modeling and seismological observation), our view is
essentially the same as that of Morel and Butler (2008).
Assuming that the solar neon abundance may be regarded as equivalent 
to that derived for B stars (8.0), we obtain the Ne/O ratio of
only $\sim 0.2$ ($10^{-0.7}$), even if the 3D-corrected (lower) 
solar oxygen abundance (8.7) is adopted, which rules out 
any chance of very high solar Ne/O for such a remedy.
Thus, the solution to this problem should be sought elsewhere
(see, e.g., Asplund et al. 2009, who concluded Ne/O~=~0.17).

\section{Conclusion}

In order to see whether any significant difference exists between 
the oxygen composition in the current galactic gas (almost equivalent 
to the photospheric material of young stars such as B-type stars) and in the solar 
atmosphere, which has been reported so far (i.e., the former being appreciably 
lower by $\sim 0.3$~dex) and regarded as a serious problem because it 
apparently contradicts the chemical evolution of the Galaxy, 
we determined non-LTE oxygen abundances for 64 mid- through 
late-B stars by using the O~{\sc i} 6156--8 lines based on high-dispersion
spectral data collected at Okayama Astrophysical Observatory, and 
compared them with the solar O abundance established in the same manner.

We then obtained $\langle A^{\rm O} \rangle = 8.71 (\pm 0.06)$ as the average
oxygen abundance of the sample B stars, resulting in the star$-$Sun 
differential abundance of $\langle$[O/H]$\rangle = -0.10$ in
comparison with $A_{\odot}^{\rm O} = 8.81$ derived by Takeda and Honda (2005),
which means that the difference is less significant than previously thought. 
Moreover, if an appropriate 3D-effect correction of $\sim -0.1$~dex is
applied to the solar abundance according to Caffau et al. (2008),
the $\langle A^{\rm O} \rangle$ for these B stars turns out to be 
almost the same as $A_{\odot}^{\rm O}$. We thus conclude that
no essential abundance discrepancy exists between the Sun and
the current galactic gas, as far as oxygen is concerned.

This does not imply, however, that the abundance consistency of
the interstellar gas and the solar photosphere similarly holds
for other elements. While the apparently subsolar trend of C or N 
in B-type stars (see, e.g., figure 1 of Nissen 1993) may admittedly 
be mitigated (as we found for oxygen in this study) when the downward 
revision of the solar abundances is invoked (Asplund et al. 2009),
its authenticity is still controversial.
Also, since the tendency of slight metallicity deficit by several 
tenths dex as compared to the Sun has been reported even for superficially 
normal B or A-type stars (e.g., Sadakane 1990; Niemczura 2003; 
Niemczura et al. 2009), we should bear in mind a possibility of some 
non-canonical composition change in the galactic gas over these several 
billion years, such as massive infall of metal-poor primordial gas 
onto the disk (see, e.g., Takeda et al. 2008).

As a by-product of oxygen abundance analysis, we also determined the 
Ne abundances for these B-type stars from Ne~{\sc i} 6143/6163 lines
and obtained $\langle A^{\rm Ne} \rangle = 8.02 (\pm 0.09)$, which is 
in agreement with other published studies such as the recent work by 
Morel and Butler (2008). On the assumption that the solar photospheric
Ne abundance is nearly the same with this result for B stars (such as 
the case of oxygen we confirmed), the solar Ne/O ratio was estimated to 
be $\sim 0.2$. This rules out a possibility of a very high solar 
Ne/O ratio as much as $\sim 0.5$, which was once suggested to
circumvent the solar model problem caused by the recently claimed
downward revision of solar CNO abundances (Asplund et al. 2009). 

\bigskip

This research has made use of the SIMBAD database, operated by
CDS, Strasbourg, France.

\clearpage

\onecolumn

\setcounter{figure}{1}
\begin{figure}
  \begin{center}
    \FigureFile(160mm,230mm){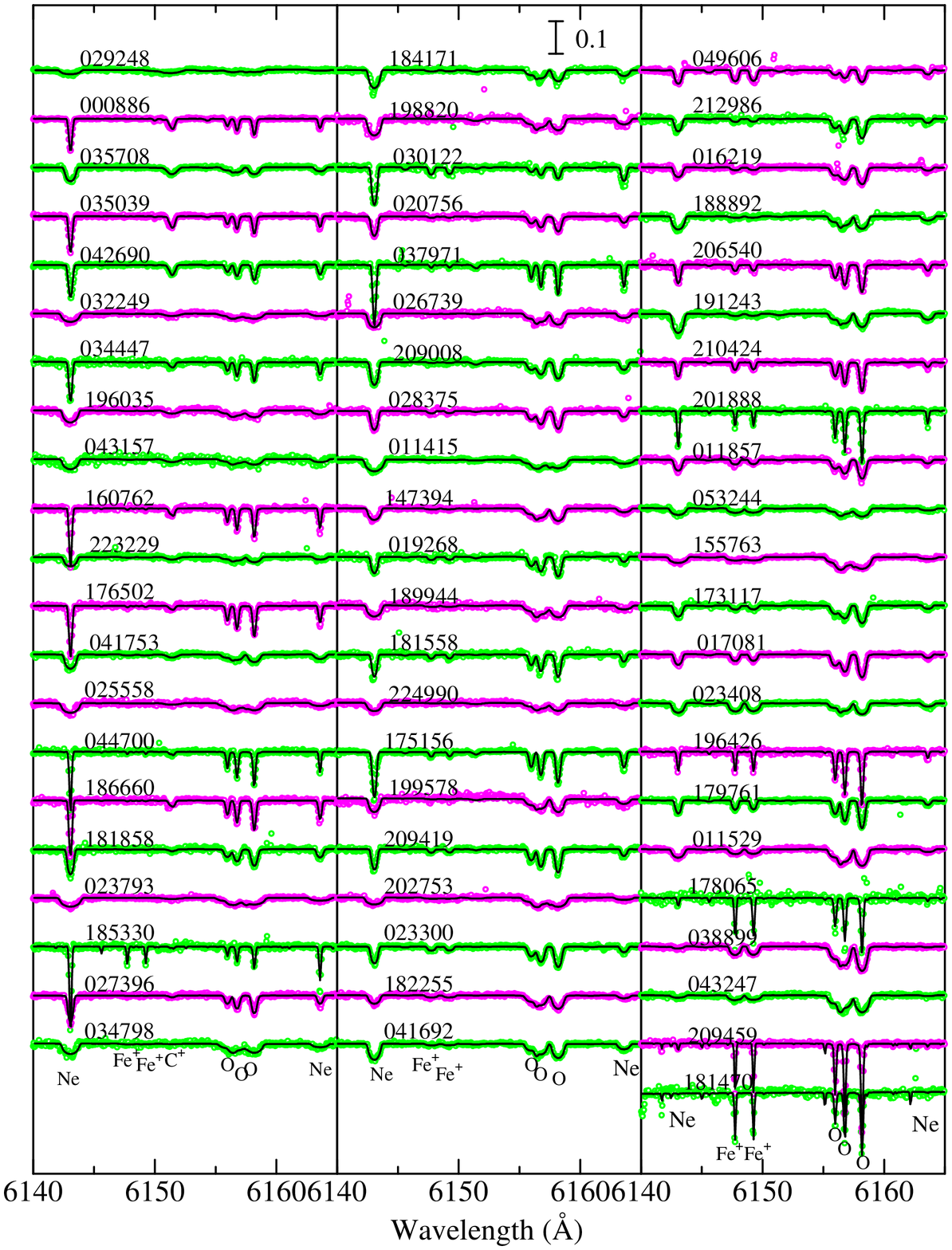}
  \end{center}
\caption{Synthetic spectrum fitting at the 6140--6165~$\rm\AA$ 
region for determining the abundances of 
O and Ne (along with C and Fe for higher-$T_{\rm eff}$ stars).
The best-fit theoretical spectra are shown by solid lines, 
while the observed data are plotted by symbols.  
In each panel (from left to right), the spectra are arranged 
(from top to bottom) in the descending order of $T_{\rm eff}$ 
as in table 1, and an offset of 0.1 is applied to each spectrum 
(indicated by the HD number) relative to the adjacent one. 
}
\end{figure}

\clearpage
\setcounter{table}{0}
\begin{table}[h]
\caption{Basic data of the target stars and the results of abundance analyses.}
\scriptsize
\begin{center}
\begin{tabular}{clcc c@{ }r@{  }c c@{ }r@{ }c@{ }r@{ }c} 
\hline\hline
HD & Sp.Type & $T_{\rm eff}$ & $\log g$ & 
$A_{\rm N}^{\rm O}$ & $W_{6156-8}^{\rm O}$ & $\Delta_{6156-8}^{\rm O}$ &
$A_{\rm N}^{\rm Ne}$ & $W_{6143}^{\rm Ne}$ & $\Delta_{6143}^{\rm Ne}$ &
$W_{6163}^{\rm Ne}$ & $\Delta_{6163}^{\rm Ne}$ \\
\hline
029248 & B2~III   & 22651& 3.58& 8.63&  23.2& $-$0.28& 7.89&  18.8& $-$0.45&   5.0& $-$0.36\\
000886 & B2~IV    & 21667& 3.83& 8.77&  37.8& $-$0.22& 8.09&  31.5& $-$0.49&   9.1& $-$0.33\\
035708 & B2.5~IV  & 21082& 4.09& 8.77&  40.1& $-$0.17& 8.13&  35.5& $-$0.47&  10.7& $-$0.29\\
035039 & B2~IV-V  & 20059& 3.69& 8.72&  42.1& $-$0.22& 8.06&  36.8& $-$0.53&  11.0& $-$0.34\\
042690 & B2~V     & 19299& 3.81& 8.71&  44.8& $-$0.20& 8.06&  39.7& $-$0.53&  12.2& $-$0.32\\
032249 & B3~V     & 18890& 4.13& 8.69&  44.3& $-$0.16& 7.95&  32.8& $-$0.43&   9.9& $-$0.28\\
034447 & B2~V     & 18480& 4.10& 8.69&  46.6& $-$0.16& 8.02&  38.7& $-$0.46&  12.0& $-$0.28\\
196035 & B3~IV    & 17499& 4.36& 8.70&  52.4& $-$0.12& 8.08&  42.3& $-$0.43&  13.8& $-$0.24\\
043157 & B5~V     & 17486& 4.12& 8.61&  44.0& $-$0.15& 7.94&  36.5& $-$0.43&  11.3& $-$0.26\\
160762 & B3~IV    & 17440& 3.91& 8.74&  59.2& $-$0.18& 8.08&  48.8& $-$0.55&  16.0& $-$0.30\\
223229 & B3~IV    & 17327& 4.20& 8.49&  35.2& $-$0.13& 7.82&  28.5& $-$0.36&   8.5& $-$0.24\\
176502 & B3~V     & 16821& 3.89& 8.78&  69.5& $-$0.19& 8.04&  46.6& $-$0.52&  15.2& $-$0.29\\
041753 & B3~V     & 16761& 3.90& 8.57&  46.1& $-$0.17& 7.94&  39.7& $-$0.47&  12.5& $-$0.28\\
025558 & B3~V     & 16707& 4.29& 8.75&  63.5& $-$0.13& 8.01&  38.4& $-$0.39&  12.3& $-$0.23\\
044700 & B3~V     & 16551& 4.21& 8.73&  63.4& $-$0.14& 8.04&  41.4& $-$0.42&  13.5& $-$0.24\\
186660 & B3~III   & 16494& 3.57& 8.73&  67.5& $-$0.23& 8.06&  53.2& $-$0.62&  17.7& $-$0.34\\
181858 & B3~IVp   & 16384& 4.19& 8.74&  65.9& $-$0.14& 8.07&  43.4& $-$0.44&  14.4& $-$0.24\\
023793 & B3~V+... & 16264& 4.15& 8.73&  65.9& $-$0.15& 8.01&  39.3& $-$0.41&  12.7& $-$0.24\\
185330 & B5~II-III& 16167& 3.77& 8.35&  31.7& $-$0.17& 8.15&  57.2& $-$0.61&  20.0& $-$0.31\\
027396 & B4~IV    & 16028& 3.91& 8.66&  60.3& $-$0.17& 7.95&  38.9& $-$0.44&  12.4& $-$0.26\\
034798 & B5~IV/V  & 15943& 4.27& 8.77&  73.1& $-$0.13& 8.05&  38.3& $-$0.38&  12.6& $-$0.21\\
184171 & B3~IV    & 15858& 3.54& 8.69&  67.3& $-$0.22& 8.01&  49.2& $-$0.57&  16.2& $-$0.32\\
198820 & B3~III   & 15852& 3.86& 8.80&  81.2& $-$0.19& 8.13&  52.6& $-$0.55&  18.2& $-$0.28\\
030122 & B5~III   & 15765& 3.72& 8.38&  35.5& $-$0.17& 8.15&  57.1& $-$0.60&  20.1& $-$0.30\\
020756 & B5~IV    & 15705& 4.43& 8.68&  62.0& $-$0.11& 8.07&  34.4& $-$0.32&  11.3& $-$0.18\\
037971 & B4/B5~III& 15532& 3.63& 8.67&  67.2& $-$0.21& 8.17&  59.1& $-$0.63&  21.0& $-$0.31\\
026739 & B5~IV    & 15490& 3.92& 8.75&  76.6& $-$0.17& 8.04&  42.4& $-$0.45&  14.0& $-$0.25\\
209008 & B3~III   & 15353& 3.50& 8.72&  75.6& $-$0.23& 8.00&  47.2& $-$0.54&  15.6& $-$0.30\\
028375 & B3~V     & 15278& 4.30& 8.77&  79.1& $-$0.12& 8.10&  35.8& $-$0.34&  11.9& $-$0.19\\
011415 & B3~III   & 15174& 3.54& 8.73&  78.5& $-$0.22& 8.05&  49.0& $-$0.54&  16.5& $-$0.29\\
147394 & B5~IV    & 14898& 4.01& 8.76&  82.6& $-$0.16& 8.03&  34.6& $-$0.36&  11.3& $-$0.21\\
019268 & B5~V     & 14866& 4.24& 8.75&  80.8& $-$0.13& 8.01&  28.6& $-$0.29&   9.2& $-$0.17\\
189944 & B4~V     & 14793& 4.01& 8.79&  89.4& $-$0.16& 8.11&  38.3& $-$0.39&  12.9& $-$0.21\\
181558 & B5~III   & 14721& 4.15& 8.71&  76.6& $-$0.14& 8.06&  31.9& $-$0.32&  10.5& $-$0.18\\
224990 & B4~III   & 14569& 3.99& 8.54&  56.2& $-$0.14& 7.95&  28.2& $-$0.32&   8.9& $-$0.19\\
175156 & B3~II    & 14561& 2.79& 8.73&  86.8& $-$0.33& 8.00&  56.1& $-$0.68&  19.0& $-$0.37\\
199578 & B5~V     & 14480& 4.02& 8.70&  78.3& $-$0.15& 8.14&  36.8& $-$0.37&  12.5& $-$0.20\\
209419 & B5~III   & 14404& 3.82& 8.73&  84.9& $-$0.18& 7.99&  32.3& $-$0.36&  10.4& $-$0.21\\
202753 & B5~V     & 14318& 3.84& 8.68&  78.5& $-$0.17& 7.91&  27.3& $-$0.32&   8.5& $-$0.20\\
023300 & B6~V     & 14207& 3.84& 8.73&  86.9& $-$0.17& 8.02&  31.6& $-$0.35&  10.3& $-$0.20\\
182255 & B6~III   & 14190& 4.29& 8.72&  82.7& $-$0.12& 8.04&  23.4& $-$0.27&   7.5& $-$0.14\\
041692 & B5~IV    & 14157& 3.19& 8.70&  87.7& $-$0.27& 8.00&  45.1& $-$0.53&  15.1& $-$0.29\\
049606 & B7~III   & 14121& 3.82& 8.41&  47.1& $-$0.15& 7.90&  25.4& $-$0.31&   7.9& $-$0.19\\
212986 & B5~III   & 14121& 4.27& 8.75&  88.6& $-$0.12& 8.08&  24.5& $-$0.29&   8.0& $-$0.14\\
016219 & B5~V     & 14113& 4.06& 8.72&  85.7& $-$0.14& 7.93&  22.1& $-$0.24&   6.9& $-$0.16\\
188892 & B5~IV    & 14008& 3.38& 8.67&  83.2& $-$0.23& 8.01&  39.5& $-$0.46&  13.1& $-$0.26\\
206540 & B5~IV    & 13981& 4.01& 8.72&  87.0& $-$0.15& 8.00&  24.9& $-$0.28&   7.9& $-$0.17\\
191243 & B5~Ib    & 13923& 2.50& 8.66&  81.3& $-$0.35& 7.95&  52.3& $-$0.67&  17.5& $-$0.38\\
210424 & B5~III   & 13740& 3.99& 8.69&  85.8& $-$0.14& 7.86&  18.0& $-$0.22&   5.5& $-$0.15\\
201888 & B7~III   & 13689& 4.01& 8.74&  93.8& $-$0.15& 8.06&  24.3& $-$0.27&   7.9& $-$0.16\\
011857 & B5~III   & 13600& 3.88& 8.69&  87.1& $-$0.16& 7.98&  22.6& $-$0.25&   7.2& $-$0.16\\
053244 & B8~II    & 13467& 3.42& 8.46&  59.8& $-$0.19& 7.83&  23.5& $-$0.30&   7.2& $-$0.21\\
155763 & B6~III   & 13397& 4.24& 8.78& 105.9& $-$0.12& 8.23&  23.6& $-$0.27&   8.0& $-$0.13\\
173117 & B8~III   & 13267& 3.63& 8.65&  87.4& $-$0.18& 7.98&  23.8& $-$0.32&   7.6& $-$0.18\\
017081 & B7~IV    & 13063& 3.72& 8.73& 104.0& $-$0.18& 8.00&  20.9& $-$0.24&   6.7& $-$0.16\\
023408 & B8~III   & 12917& 3.36& 8.49&  68.6& $-$0.19& 8.06&  28.5& $-$0.35&   9.5& $-$0.20\\
196426 & B8~IIIp  & 12899& 3.89& 8.73& 105.0& $-$0.15& 7.89&  13.9& $-$0.19&   4.2& $-$0.13\\
179761 & B8~II-III& 12895& 3.46& 8.69&  99.9& $-$0.20& 7.92&  20.7& $-$0.26&   6.5& $-$0.18\\
011529 & B8~III   & 12858& 3.43& 8.74& 109.5& $-$0.22& 8.01&  24.5& $-$0.31&   8.0& $-$0.19\\
178065 & B9~III   & 12243& 3.49& 8.60&  94.1& $-$0.18& 7.38&   5.1& $-$0.16&   1.4& $-$0.13\\
038899 & B9~IV    & 10774& 4.02& 8.73& 145.4& $-$0.10& 8.26&   5.3& $-$0.09&   1.6& $-$0.06\\
043247 & B9~II-III& 10301& 2.39& 8.55& 123.1& $-$0.23& 7.80&   8.0& $-$0.19&   2.4& $-$0.14\\
209459 & B9.5~V   & 10204& 3.53& 8.66& 142.5& $-$0.13& 8.08&   3.6& $-$0.10&   1.1& $-$0.07\\
181470 & A0~III   & 10085& 3.92& 8.44&  95.8& $-$0.08& 7.41&   0.5& $-$0.05&   0.1& $-$0.04\\
\hline
\end{tabular}
\end{center}
In columns 1 through 4 are given the HD number, the spectral type (from 
SIMBAD database), the effective temperature (in K), and the logarithmic 
surface gravity (in  cm~s$^{-2}$). Columns 5--7 and 8--12 show the 
results of the abundance analysis for O and Ne, respectively.
That is, $A_{\rm N}$ is the non-LTE abundance (in the usual normalization
of H = 12.00) derived from the spectrum-synthesis fitting, $W$ is the 
equivalent width (in m$\rm\AA$) for the total O~6156--8 triplet 
(comprising 9 components; cf. table 2) or Ne~6143 or Ne~6163 inversely 
computed from $A_{\rm N}$, and $\Delta$ is the non-LTE correction 
($\equiv A_{\rm N} - A_{\rm L}$). 
The 64 stars are arranged in the descending order of $T_{\rm eff}$. 
\end{table}

\clearpage
\setcounter{table}{1}
\begin{table}[h]
\caption{Adopted atomic data of O~{\sc i} and Ne~{\sc i} lines.}
\begin{center}
\begin{tabular}
{cccccc}
\hline \hline
Desig. & Species & transition & $\lambda (\rm\AA) $ & $\chi_{\rm low}$~(eV)  & $\log gf$ \\
\hline
O~6156 & O~{\sc i} & 3p$^{5}P_{1}$--4d$^{5}D_{0}$ &  6155.961 & 10.74 & $-1.40$ \\
       & O~{\sc i} & 3p$^{5}P_{1}$--4d$^{5}D_{1}$ &  6155.971 & 10.74 & $-1.05$ \\
       & O~{\sc i} & 3p$^{5}P_{1}$--4d$^{5}D_{2}$ &  6155.989 & 10.74 & $-1.16$ \\
O~6157 & O~{\sc i} & 3p$^{5}P_{2}$--4d$^{5}D_{1}$ &  6156.737 & 10.74 & $-1.52$ \\
       & O~{\sc i} & 3p$^{5}P_{2}$--4d$^{5}D_{2}$ &  6156.755 & 10.74 & $-0.93$ \\
       & O~{\sc i} & 3p$^{5}P_{2}$--4d$^{5}D_{3}$ &  6156.778 & 10.74 & $-0.73$ \\
O~6158 & O~{\sc i} & 3p$^{5}P_{3}$--4d$^{5}D_{2}$ &  6158.149 & 10.74 & $-1.89$ \\
       & O~{\sc i} & 3p$^{5}P_{3}$--4d$^{5}D_{3}$ &  6158.172 & 10.74 & $-1.03$ \\
       & O~{\sc i} & 3p$^{5}P_{3}$--4d$^{5}D_{4}$ &  6158.187 & 10.74 & $-0.44$ \\
\hline
Ne~6143& Ne~{\sc i} & 3s[3/2]$_{2}$--3p[3/2]$_{2}$ & 6143.063 & 16.62 & $-0.07$ \\
Ne~6163& Ne~{\sc i} & 3s[1/2]$_{0}$--3p[1/2]$_{1}$ & 6163.594 & 16.72 & $-0.61$ \\
\hline
\end{tabular}
\end{center}
The atomic data of these O~{\sc i} lines (multiplet 10) were adopted from
Kurucz and Bell's (1995) compilation, which are based on the NBS values
(Wiese et al. 1966), while those for the Ne~{\sc i} lines were taken
from Morel and Butler (2008).
\end{table}

\end{document}